\documentclass[useAMS, usenatbib]{mn2e}
\usepackage{graphics,epsfig,psfig}
\usepackage[normalem]{ulem}
\usepackage{xcolor}
\usepackage{amsmath, amssymb}
\usepackage[]{inputenc,amssymb}
\usepackage{graphicx}

\def \msun{M$_\odot \,$ }
\def \mpy{\msun yr$^{-1}\,$ }

\def \mpcc{m$_p$ cm$^{-3}$ }
\def \ergps{erg s$^{-1}\,$}
\def \kmps{km s$^{-1}\,$}
\def \etal{{et al.\ }}

\usepackage[pdfpagelabels]{hyperref}
\hypersetup{%
   colorlinks=true,hyperfootnotes=false,%
   breaklinks=true,%
   plainpages=false, bookmarksnumbered, bookmarksopen=true,
   bookmarksopenlevel=1,%
   urlcolor=blue, linkcolor=blue, citecolor=blue,
   }

\title[Fermi Bubbles: clues from OVIII/OVII line ratio]{Clues to the origin of Fermi Bubbles from OVIII/OVII line ratio}

\author[Sarkar, Nath and Sharma]
{Kartick C. Sarkar$^{1,2}$ \thanks{kcsarkar@rri.res.in}, Biman B. Nath$^1$ \thanks{biman@rri.res.in}, Prateek Sharma$^2$\\
$^1$Raman Research Institute, Sadashiva Nagar, Bangalore 560080, India\\
$^2$Joint Astronomy Programme and Department of Physics, Indian Institute of Science, Bangalore 560012, India\\
}

\begin{document}

\maketitle

\label{firstpage}

\begin{abstract}
We constrain the origin of Fermi Bubbles using 2D hydrodynamical simulations of both star formation driven  and black hole accretion driven wind models. We compare our results with recent observations of OVIII to OVII line ratio within and near Fermi Bubbles. Our results suggest that independent of the driving mechanisms,  a low luminosity ($\mathcal{L} \sim 0.7-1\times 10^{41}$ erg s$^{-1}$) energy injection best reproduces the observed line ratio for which the shock temperature is $\approx 3\times 10^6$ K. Assuming the Galactic halo temperature to be $2\times 10^6$K, we estimate the shock velocity to be $\sim 300$ km s$^{-1}$ for a weak shock. The corresponding estimated age of the Fermi bubbles is $\sim 15-25$ Myr. Such an event can be produced either by a star formation rate of $\sim 0.5$ M$_\odot$ yr$^{-1}$ at the Galactic centre or a very low luminosity jet/accretion wind arising from the central black hole. Our analysis rules out any activity that generates an average mechanical luminosity $\gtrsim 10^{41}$ \ergps as a possible origin of the Fermi Bubbles.
\end{abstract}


\section{Introduction}
\label{sec:intro}
The discovery of Fermi Bubbles (FBs) \citep{su2010, ackermann2014} has given a boost for studying the interaction of Galactic wind and the circum-galactic medium (CGM) of the Milky Way (MW). They are also excellent laboratories to study high energy astrophysics phenomena in such systems as they produce radiation ranging from radio to gamma-rays. However, the dynamical and spectral origin of these bubbles still remain debatable even after $\sim 6$ years of their discovery. 

The dynamical models of the FBs can be divided mainly into two categories. First, AGN driven models in which the bubbles originate from a past accretion activity of the MW central black hole over a time scale of $3\hbox{--}12$ Myr, with a  luminosity of $\sim 2\times 10^{41}\hbox{--} 10^{43}$ \ergps, either via an accretion wind (AGNW) \citep{zubovas2011, zubovas2012, mou2014, mou2015} or via a jet \citep{guo2012, yang2012} from the Galactic centre black hole. Second, a star formation (SF) driven wind model (SFW) in which the bubbles originate from supernovae activity due to star formation at the centre of our Galaxy \citep{lacki2014, crocker2014b, sarkar2015b}. Based on the star formation rate (SFR) at the centre, the age of the bubbles has been estimated to range from $\sim 25$ Myr \citep[hereafter, S15]{sarkar2015b} to $\sim 200$ Myr \citep{crocker2014b}.  

Although the observations suggest that the current accretion rate of the Galactic centre black hole (GCBH) is $\sim 10^{-9}\,\hbox{--}\, 10^{-7}$ \mpy \citep{quataert2000, agol2000, yuan2003, marrone2006} corresponding to a mechanical luminosity of $\sim 5\times 10^{36\hbox{--}38}$ \ergps, it has been suggested that it could have been several orders of magnitude higher in the past \citep{totani2006}. On the other hand, infra-red observations suggest that the current SFR is $\approx 0.1$ \mpy \citep{yusuf-zadeh2009}, compared to the rate of $\approx 0.3$ \mpy required to produce the bubbles \citep{sarkar2015b}.

The spectral models of the FBs can also be divided into mainly two types. First, the hadronic models, in which the gamma-rays are emitted via interactions between cosmic ray (CR) protons and gas phase protons \citep{crocker2011, crocker2012, crocker2014b, mou2014, mou2015}. Second, the leptonic models, in which low energy photons (either cosmic microwave background or interstellar radiation field) are energised \textit{in situ} by high energy cosmic ray electrons to produce gamma rays \citep{su2010, mertsch2011, sarkar2015b}. 

While modelling the gamma rays requires knowledge of the local cosmic ray (CR) energy density, magnetic field and gas density, and involves some assumptions about the acceleration processes and diffusion of the CRs, the modelling of the bubbles is much simpler in X-rays as it involves only the local gas density and its temperature. From the lack of X-ray emission inside the bubbles it has been suggested that these bubbles are under-dense compared to the surroundings. However, measuring the  density inside and outside the bubbles requires a careful fitting of the emission or absorption spectra. 

An ideal place to measure the spectra would be the northern polar spur (NPS) where the shell is X-ray bright. However, there have been debates over the actual distance of the NPS. Early observations suggested that the NPS can be a nearby low density bubble created by the stellar wind from the Scorpio-Centaurus OB association or could be a supernova remnant situated at a distance of a few hundred pc \citep{berkhuijsen1977}.  Using X-ray observations \cite{sofue1994, snowden1995, lallement2016}, however, found that the NPS feature is heavily absorbed towards the Galactic plane requiring a hydrogen column density of $\sim$ few $\times 10^{21}$ cm$^{-2}$ which makes it unlikely to be a nearby feature. Recent observations using \textit{Suzaku} and \textit{XMM-Newton} also found  that the spectra can be better explained if the NPS feature is of the `Galactic centre origin' (see section 4.3 of \cite{kataoka2013} for a detailed discussion). Another recent observation of OVIII Ly-$\alpha$ to Ly-$\beta$ ratio by \cite{gu2016} also supported the `Galactic centre origin' of the NPS \citep{sofue1977, sofue2000, blandhawthorn2003, sarkar2015b, sofue2016}. Also it would be a  \textit{dramatic coincidence} that the inner edge of the NPS traces the outer edge of the FBs even at high latitudes if the NPS is not related to the FBs. 

Individual pointings towards NPS, therefore, have been used several times to estimate the post shock temperature of the FBs. Observations by \cite{snowden1995, kataoka2013, gu2016} suggested that the temperature of the NPS is $\sim 0.25-0.3$ keV corresponding to a Mach number ($\mathcal{M}$) of $\sim 1.5$, considering the halo temperature $\approx 2\times 10^6$K (estimated from the OVIII to OVII line ratio \citep{miller2015}). Not only at the NPS, absorption study of OVII lines towards 3C 273, $\approx (-60^\circ,+60^\circ)$, also suggests a shock velocity of $\approx 200\hbox{--}300$ \kmps \citep{fang2014}. These suggest a star formation driven or a low luminosity AGNW driven origin for the FBs (since the stronger AGNW would produce a stronger shock with $\mathcal{M} \gg 1$). However, in a recent observation of the OVIII to OVII line intensity ratio \citealt[(hereafter, MB16)]{miller2016} found that the sight-lines passing through FBs and the surroundings (except the NPS) have a temperature $\approx 5\times 10^6$ K. This led them to conclude that the shock is because of an AGN activity at the Galactic centre and the age of the FBs is $\sim 4$ Myr (see also \cite{nicastro2016}). This differs from other estimates of a lower temperature and a weaker shock, and therefore a longer age of the FBs.

In this paper, we perform 2D hydrodynamical simulations of both star formation driven and accretion wind driven bubbles in a realistic MW gravity and a self consistent halo gas which is also close to the observed density distribution. We generate projected OVIII to OVII line intensity maps and ratio towards the FBs for a range of injected luminosities and compare them with the observations of MB16. Based on our simulated intensity maps, we constrain the age of the FBs and the strength of the star formation or the accretion wind/ jet activity at the Galactic centre. We also discuss the effects of conduction and the electron-proton equilibration time-scale on our results.

The paper is organised as follows. Section \ref{sec:halo} discusses the choice of dark matter and disc potential, and the hydrostatic hot halo gas. The simulation details and other code parameters are explained in section \ref{sec:sim_details}. In section \ref{sec:tool} we discuss the tools for projecting our 2D simulation results into a surface brightness map of OVIII to OVII line ratio at the Solar location. We present our results in section \ref{sec:results} and finally discuss the implications of the results in section \ref{sec:discussion}. 

\section{Galactic halo distribution}
\label{sec:halo}
One issue while modelling the FBs is the density and temperature distribution of the Galactic halo gas which carries crucial information about the soft X-ray background and also determines the shape and speed of any shock travelling through it. Because of our off-centred location, which is $\sim 8.5$ kpc away from the Galactic centre, it is in principle possible to determine the density distribution of the halo. However, there is a split in the opinion as to the correct density distribution. Based on ram pressure stripping of the dwarf satellites, the density has been estimated to be $\sim 1.3\hbox{--}3.8\times 10^{-4}$ \mpcc within $50\hbox{--}90$ kpc \citep{gatto2013}, whereas, based on the distribution of the OVII and OVIII lines, \cite{miller2015} find 
\begin{equation}
\label{eq:mb15-halo}
n(r) = n_0\,\left(1+(r/r_c)^2 \right)^{-3\beta/2}
\end{equation}
with $\beta = 0.5$, $n_0 r_c^{3\beta} = 1.35\times 10^{-2}$ cm$^{-3}$ kpc$^{3\beta}$ and $r_c < 5$ kpc, which predicts a higher density at the same distance range. A probable solution is that the metallicity of the halo is gradually decreasing with radius. Therefore, a higher density is not apparent in OVII or OVIII line emission \citep{troitsky2016}. However, there is much to be worked out before making any firm conclusion.

\begin{figure}
\centering
\includegraphics[trim={1cm 0 0 0},clip=true,width=0.4\textheight]{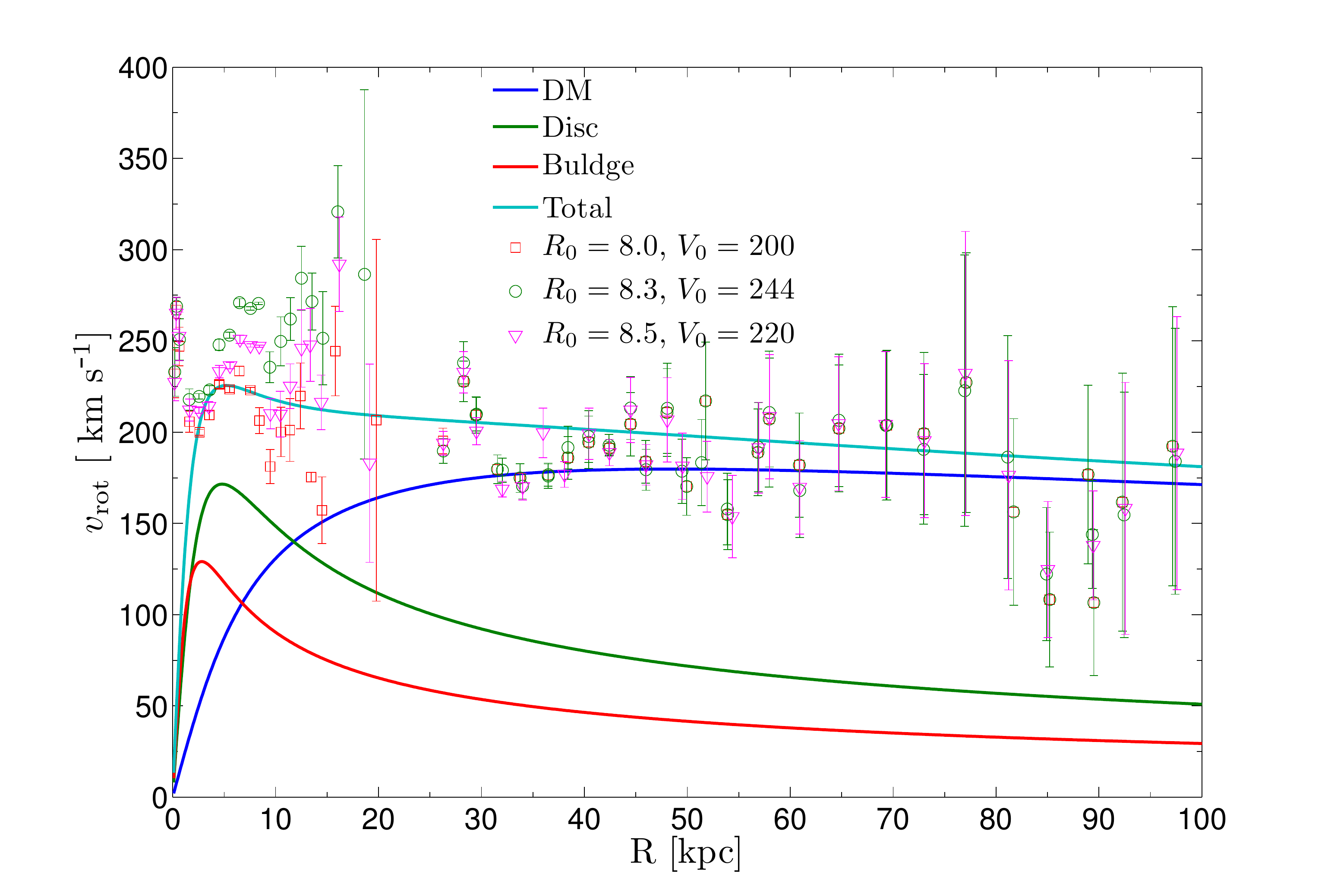}
\caption{Rotation curve for the assumed gravitational fields for the parameters given in table \ref{table:halo-param}. Data points from \protect \cite{bhattacharjee2013} have been shown with the errorbars. Different color of the data points represent assumed Solar distance from the Galactic centre ($R_0$ in kpc) and Solar rotation velocity ($V_0$ in \kmps).} 
\label{fig:vrot}
\end{figure}
\begin{figure}
\centering
\includegraphics[trim={6cm 0 0 0},clip=true, width=0.4\textheight]{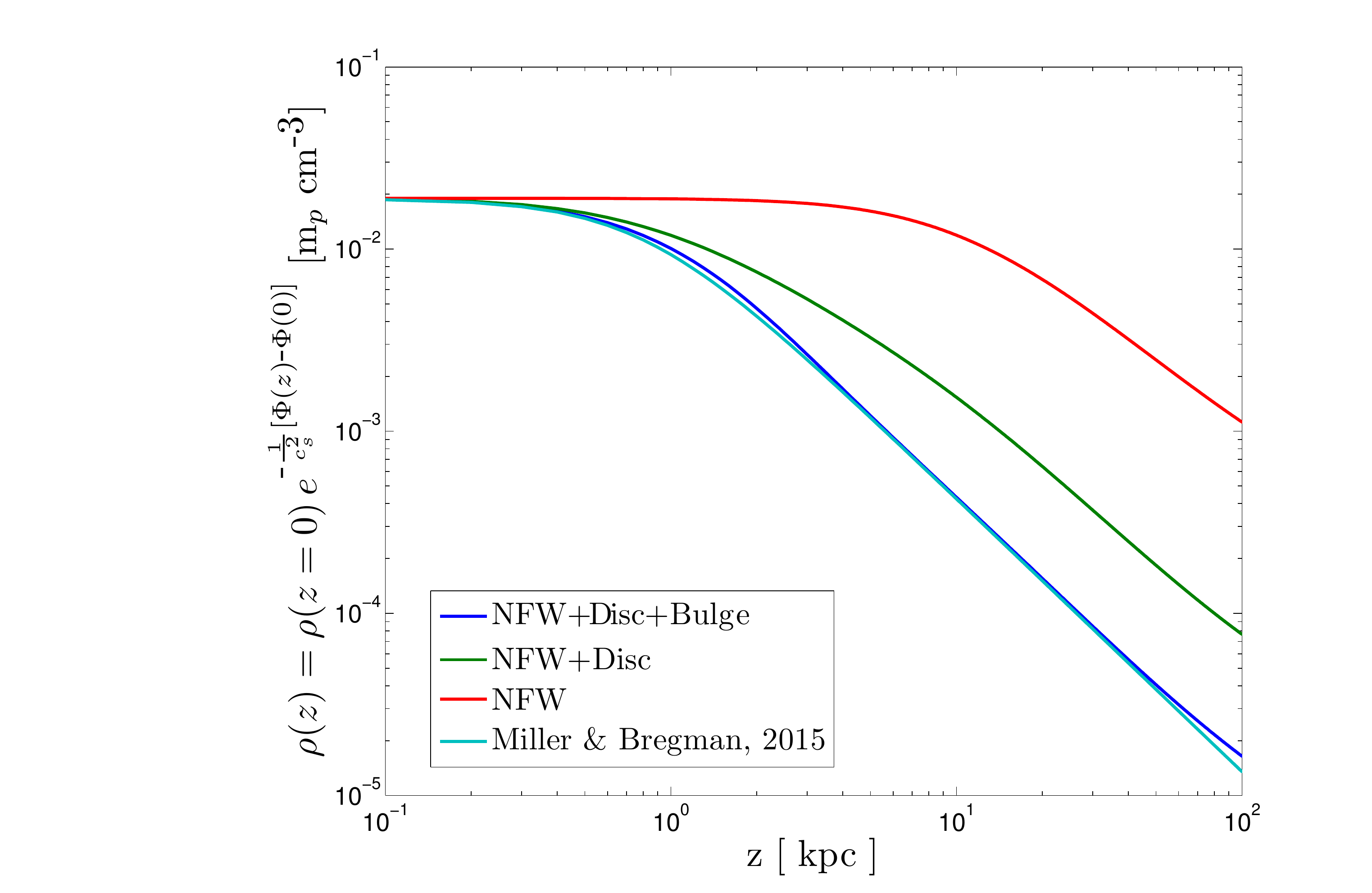}
\caption{Equilibrium density distribution of the halo gas (Eq. \ref{eq:eq-halo}) for the parameters given in table \ref{table:halo-param}. The red curve shows the distribution in case of only NFW potential, green curve shows the distribution if the stellar disc is added, blue curve shows the distribution once all the components have been added together. The cyan curve shows the best fitting halo distribution from \protect \cite{miller2015}.} 
\label{fig:rho}
\end{figure}

In this paper we assume that the hot halo gas (isothermal at temperature, $T_{\rm halo} = 2\times 10^6$ K) is in hydrostatic equilibrium with the gravity of the dark matter, the disc stars and the bulge. For the dark matter, we use NFW gravity \citep{nfw1996}, with an added core to ensure finite dark matter density at $r = 0$,
\begin{equation}
\Phi_{\rm DM} = -\frac{GM_{\rm vir}}{f(c)} \frac{\log\left(1+\sqrt{r^2+d^2}/r_s\right)}{\sqrt{r^2+d^2}}\,.
\end{equation}
Here, $M_{\rm vir}$ is the dark matter mass, $f(c) = \log(1+c)-c/(1+c)$ with $c$ as the concentration parameter of the dark matter distribution, $r_s$ is the scale radius, $d$ is the core radius. For the disc gravity, we use the Miyamoto \& Nagai potential \citep{miyamoto1975}
\begin{equation}
\Phi_{\rm disc} = - \frac{G M_{\rm disc}}{\sqrt{R^2+\left( a+\sqrt{z^2+b^2}\right)^2}}\,,
\end{equation}
where, $M_{\rm disc}$ is the disc mass, $R$ and $z$ are, respectively, the cylindrical radius and height, $a$ and $b$ represent the scale radius and scale height for the disc gravity. To make the gravity realistic near the Galactic centre, we also add a bulge potential of the form
\begin{equation}
\Phi_{\rm bulge} = -\frac{G M_{\rm bulge}}{\sqrt{r^2+a_b^2}}\,,
\end{equation}
where, $a_b$ is the scale radius for the bulge.

The hydrostatic density distribution for the combined gravity, $\Phi = \Phi_{\rm DM} + \Phi_{\rm disc} +\Phi_{\rm bulge}$ can, therefore, be written as 
\begin{equation}
\label{eq:eq-halo}
\rho(R,z) = \rho(0,0)\, \exp\left(-\frac{1}{c_s^2} \left( \Phi(R,z)-\Phi(0,0)\right)\right)\,,
\end{equation}
where, $\rho(0,0)$ is the density at $r = 0$ and $c_s = \sqrt{k_BT/\mu m_p}$ is the isothermal sound speed at temperature $T$ (for a detailed discussion, see \cite{sarkar2015a}).  However, note that unlike S15, we do not use a rotating cold disc component as our focus is to study the interaction of the wind and the halo gas, in particular the outer shock properties. The interaction of the wind with the disc gas affects the formation of cold clumps. These cold clumps, however, will not affect the observed OVII and OVIII properties.
 \begin{table}
  \centering
  \begin{tabular}{ l c l} 
   \hline\hline 
  parameters & values\\[0.5ex]
   \hline
   $M_{\rm vir} ({\rm M}_{\odot}) $                & $1.2\times 10^{12}$\\ 
   $M_{\rm disc } ({\rm M}_{\odot})$             & $6 \times 10^{10}$ \\
    $M_{\rm bulge } ({\rm M}_{\odot})$           & $2 \times 10^{10}$ \\
   $T_{\rm halo}$ (K)  		                               & $2 \times 10^6$ \\
   $c$ 														    & $12$  \\ 
   $r_s $ (kpc)                                                   & $21.5$ \\
   $a$ (kpc)                                                       & $3.0$ \\
   $b$ (kpc)                                                        & $0.4$ \\ 
   $d$ (kpc)                                                       & $6.0$ \\
   $\rho_{c}(0,0)$ (m$_p$cm$^{-3}$)           & $1.9\times 10^{-2}$\\
   \hline   
  \end{tabular}
     \caption{Parameters used for the mass model of our Galaxy. The assumed mass for different components are roughly consistent with the measurements by \protect \cite{mcmillan2011, mcmillan2016}.}
      \label{table:halo-param}
 \end{table}

Figure \ref{fig:vrot} shows the rotation velocity on $z=0$ plane for the parameters given in table \ref{table:halo-param}. For comparison with the observations, data from \cite{bhattacharjee2013} are shown in the same figure.  It shows an excellent consistency with the observed rotation curve of the Galaxy. 

The gas density distribution that is in hydrostatic equilibrium with the given gravity (Eq. \ref{eq:eq-halo}) is shown in figure \ref{fig:rho}. The figure also shows the effects of adding all the gravity components together.  In fact, for the given parameters, the equilibrium density distribution shows an excellent match with the standard $\beta$-model obtained by \cite{miller2015} (equation \ref{eq:mb15-halo} and shown by the cyan line in figure \ref{fig:rho}) with $\beta = 0.5$ and $r_c = 0.8$ kpc. Therefore, the hydrostatic equilibrium of MW halo gas distribution can be naturally explained by the total gravitational fields of the MW. 

\begin{figure*}
\centering
\includegraphics[width=0.4\textheight, angle=-90]{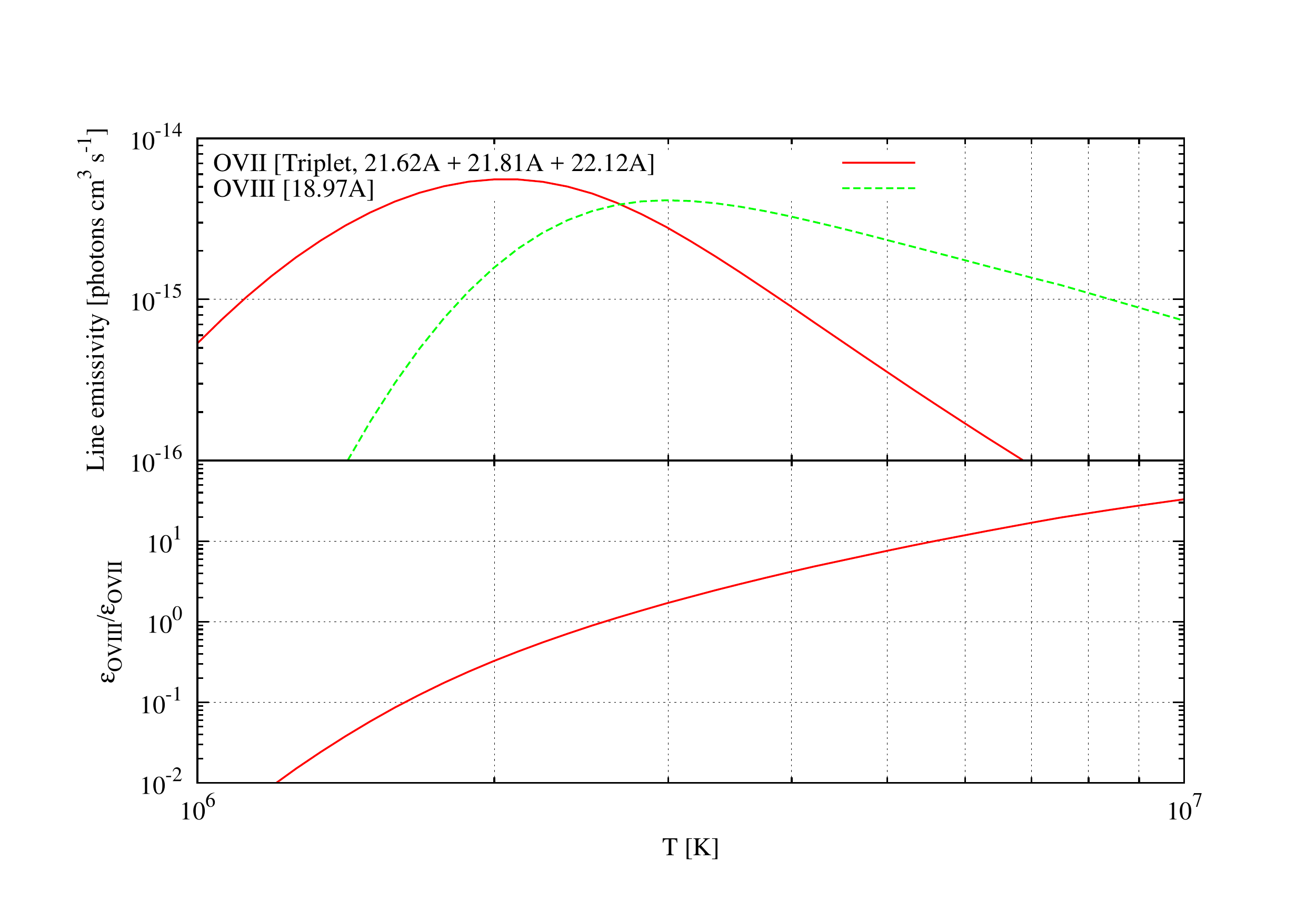}
\caption{Top panel: Temperature dependence of emissivities of OVII and OVIII lines in units of photons s$^{-1}$ cm$^{3}$ from CLOUDY-C13.04. Bottom panel: Temperature variation of the OVIII to OVII line ratio. The emissivities obtained here are for per unit hydrogen density.} 
\label{fig:oxy-lines}
\end{figure*}
\section{Simulation details}
\label{sec:sim_details}
The simulations have been performed in 2-dimensional spherical coordinates using Eulerian grid code PLUTO-v4.0 \citep{mignone2007}. The computation box extends till $15$ kpc in the radial direction and from $0$ to ${\rm \pi}/2$ in the $\theta$-direction.  The box has been divided into $256\times 256$ grid cells with uniform grid spacing in both the directions. The inner boundary of the radial direction has been set initially to the  static distribution  and the outer boundary condition is set to outflow. Both  the $\theta$-boundaries have been set to reflective type. 

While the injection of AGN and stellar mechanical energy into the ISM differ in detail, we use simplified models for them, roughly valid at the scales of the CGM. While stellar feedback has a lower velocity and is injected at a larger scale ($\sim 100$ pc), AGN wind velocity is much faster and the injection radius is smaller ($\sim10$ pc). The essential difference between the two is that the latter have a smaller mass loading and higher velocity. The two broad classes of the models discussed here can therefore be termed as SF/low-velocity wind model and AGN/high-velocity wind model.

\subsection{Star formation driven wind (SFW)}
\label{subsec:sne_driven_wind}
In case of the star formation driven winds, the energy at the base of the wind is mostly thermal and gets deposited into $\sim100$ pc region. However, in the presence of the interstellar medium (ISM) disc, the outgoing wind gets collimated and forms a biconical shape. The amount of collimation depends on the rate at which the energy is being injected from the SNe and the density and pressure structure of the ISM. Understanding the detailed structure of this component in the central region, as it was at the time of launching the winds, is difficult to do. Since our aim is to study the outer shock strength for a range of mechanical luminosities, we avoid these issues and consider that the wind has been somehow collimated by the ISM. Therefore, we inject SNe energy at the inner boundary within some opening angle. We also tune the opening angle of the energy and mass injection for each case to roughly match the shape of the contact discontinuity with the FBs. The opening angles for individual runs and other information is provided in table \ref{table:run-param}.

The inner boundary in this case is chosen to be at $r_{\rm ej}=100$ pc, which is also the point where we inject the SNe ejected mass and energy. This radius is assumed to be the transonic point of the wind\footnote{The choice of this radius is not crucial for the results presented here. However, for the sake of completeness, we compare our results for different injection radii, i.e. different transonic radii in appendix \ref{app:injection_radii}}. Therefore, the velocity at the base ($v_{\rm ej}$) is kept half of the free wind velocity ($\approx 1000 \sqrt{\alpha/\eta}$ \kmps) \citep{cc85}. Here, $\alpha = 0.3$ is the assumed heating efficiency of the supernovae (SNe) and $\eta = 0.3$ is the mass loading factor from stellar feedback \citep{leitherer1999}. The pressure is set to be $p = \rho_{\rm ej}v_{ej}^2/\gamma$, where $\gamma = 5/3$ is the adiabatic index and $\rho_{\rm ej}$ is the density at the base. The mechanical luminosity and the mass injection rate in this case can be written in terms of $\alpha$ and $\eta$ as
\begin{equation}
\label{eq:mech-L}
\mathcal{L} \approx 3 \times 10^{41}\,\, \alpha \,\,\mbox{SFR}_{\mbox{\mpy}} \,\,\,\, \mbox{\ergps} 
\end{equation}
and
\begin{equation}
\dot{M}_{\rm inj} = \eta \,\,\mbox{SFR} \,,
\end{equation}
respectively.  Therefore, density at the base can be written as
\begin{equation}
\rho_{\rm ej} = \frac{\mathcal{L}}{2\,\Omega\, r_{\rm ej}^2\, v_{\rm ej}^3}\,,
\end{equation}
where, $\Omega$ is the solid angle within which the mass and energy are injected.

\subsection{Accretion wind (AGNW)}
\label{subsec:accretion_wind}
For AGN feedback, the spherical accretion wind is likely to be collimated by the presence of the central molecular zone (CMZ) which is extended till $\sim 250$ pc in radial direction and $\sim 50$ pc in vertical direction. Following \cite{zubovas2011, mou2014}, we model the CMZ to be a disc-like structure on the $z=0$ plane having inner radius of $80$ pc and outer radius of $240$ pc. The height to radius ratio (H/R) for the CMZ is set to be H/R$=0.25$. We have also checked for H/R$=0.15$, but the results are not affected by this change. The CMZ is in local pressure balance with the hot halo and is rotationally supported by its azimuthal velocity $v_\phi = \sqrt{R\frac{d}{dR}\Phi(R,0)}$. The density of the CMZ has been kept constant at $50$ \mpcc which means that the total CMZ mass considered is $\sim 10^8$ M$_\odot$, close to the observed value. The CMZ in our set up admittedly is not in exact equilibrium with the surroundings because of unbalanced forces in z-direction. The current set up, however, is able to hold up the CMZ in its original position for more than $40$ Myr.

The wind for this case has been launched spherically at $r_{\rm ej} = 20$ pc with a velocity $v_{\rm ej} = 0.05c$, where $c$ is the speed of light in vacuum. The wind is considered to be dominated by kinetic energy and therefore, the density at the base is set to be $\rho_{\rm ej} = 2\mathcal{L}/\Omega r_{\rm ej}^2v_{\rm ej}^3$ for a mechanical luminosity of $\mathcal{L}$.
 \begin{table}
  \centering
  \caption{List of runs and the parameters used in these runs. }
  \begin{tabular}{ l c c c} 
   \hline\hline 
   Name & Type & Luminosity & Half opening angle \\
    &  & (\ergps) &  \\[1ex]
   \hline
   S5e40   & SFW & $5\times 10^{40}$ & $45^{\circ}$ \\
   S7e40   & SFW & $7\times 10^{40}$ & $45^{\circ}$ \\
   S1e41   & SFW & $ 10^{41}$ & $45^{\circ}$ \\
   A5e40  & AGNW & $5\times 10^{40}$ & $180^{\circ}$ \\
   A1e41  & AGNW & $10^{41}$ & $180^{\circ}$ \\
   A1e42  & AGNW & $10^{42}$ & $180^{\circ}$\\
   \hline   
  \end{tabular}
    \label{table:run-param}
 \end{table}
\section{Analysis Tools}
\label{sec:tool}

\subsection{Projection tool}
\label{subsec:pass}
Since we are at the Solar position, $\approx 8.5$ kpc away from the Galactic centre, which is roughly comparable to the height ($\sim 10$ kpc) and width ($\sim 4$ kpc) of FBs, the projection effects are important. A special purpose code, called Projection Analysis Software for Simulations (PASS)
\footnote{PASS is made public and is available for download at \url{ http://www.rri.res.in/~kcsarkar/pages/about_me/codes.html }},
 has been written to project the 2D simulation data to a viewer from the Solar location (assuming axisymmetry). It can also project an 1D profile into a 2D sky map, assuming spherical symmetry of the profile. The surface brightness along any line of sight ($l,b$) is calculated as 
\begin{equation}
\label{eq:los_integration}
I(l,b) = \frac{1}{4 {\rm \pi}}\int_{\rm los} n^2 \varepsilon(T) dx \,\,\,\,\,\,\,\,\, \mbox{\ergps cm}^{-2} \mbox{Sr}^{-1},
\end{equation}
where, $n$ is the particle density and $\varepsilon (T)$ is the emissivity (\ergps cm$^{3}$) at any local point along the line of sight (LOS). It can also produce mock X-ray spectra along different LOSs assuming plasma emission code MEKAL (Mewe-Kaastra-Liedahl). Since our simulation box only extends till $15$ kpc, to produce realistic emission maps, we consider the density distribution extending till $250$ kpc, and include a local bubble centred at Sun with a radius of $200$ pc, density of $4\times10^{-3}$ \mpcc and a temperature of $= 1.2\times 10^6$ K following MB16. 

\subsection{Oxygen emission lines}
\label{subsec:oxy-lines}
We assume that the plasma is in collisional ionisation equilibrium at all temperatures $ \gtrsim 10^4$ K. We can therefore obtain the density of different ionisation levels given the metallicity and temperature. The line intensity for any species $X$ can be obtained by assuming the total gaseous number density $n$ and emissivity $\varepsilon (X,T)$. In the present case we will consider only two of the ionisation levels of oxygen, \textit{viz.} OVII and OVIII among many other ionisation species present in the medium at that temperature. Therefore, the emissivities considered here will be $\varepsilon_{\rm OVII}$ and $\varepsilon_{\rm OVIII}$. These emissivities have been obtained from CLOUDY-C13.04 \citep{ferland2013} and are described in figure \ref{fig:oxy-lines}. It is clear from the figure that the OVIII to OVII line ratio is sensitive to the temperature, making it a very useful for temperature diagnostics in the range of $10^6\hbox{--}10^7$K. 
\begin{figure*}
\centering
\includegraphics[trim={0cm 0 0 2cm},clip=true, width=0.8\textheight]{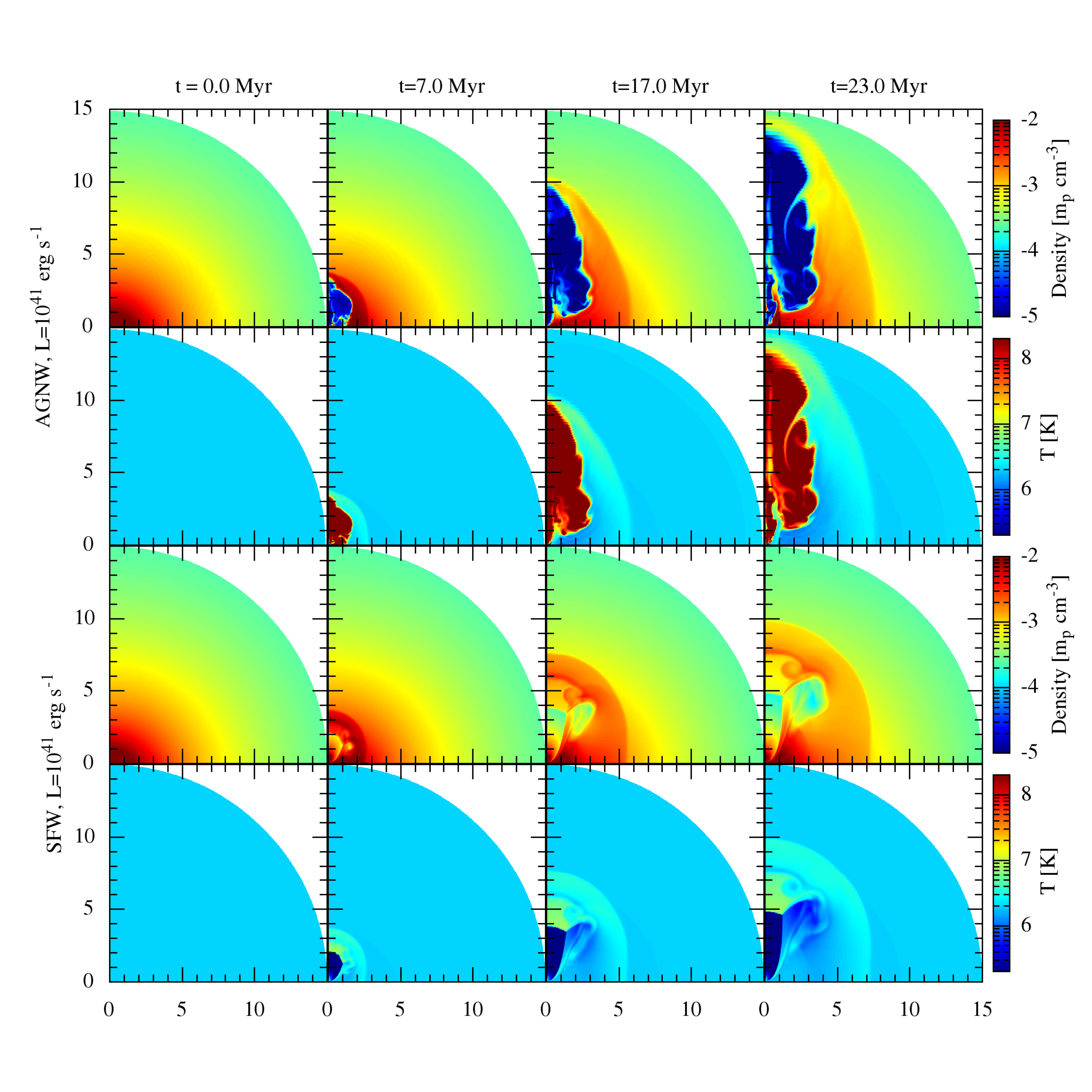}
\caption{Evolution of density and temperature contours for AGNW and SFW cases. Top two panels represent evolution of an accretion driven wind corresponding to luminosity $\mathcal{L}= 10^{41}$ \ergps, whereas, bottom two panels represent the evolution of a SF driven wind of luminosity $\mathcal{L} = 10^{41}$ \ergps (SFR = $1$ \mpy). The X-axis represents the on-plane distance $R$ [kpc], and the Y-axis represents the vertical distance, $z$ [kpc], from the Galactic disc.} 
\label{fig:rhot}
\end{figure*}
\begin{figure*}
\centering
\includegraphics[width=0.8\textheight]{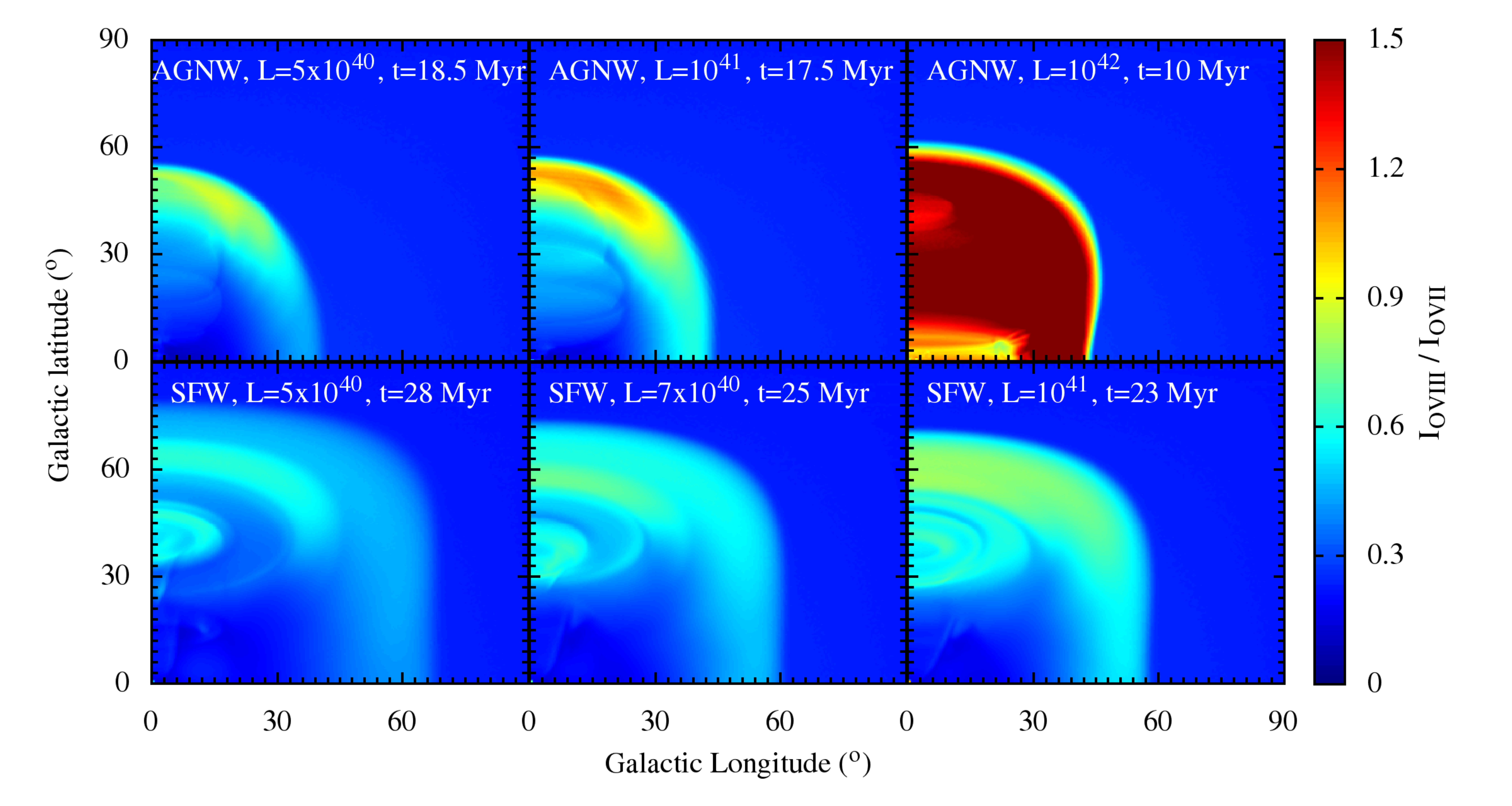}
\caption{Simulated OVIII to OVII line ratio map for all the runs mentioned in table \ref{table:run-param}. The upper panel shows the ratio for AGNW cases and the lower panel shows for SFW cases. The maps are obtained at $t=t_{\rm fb}$, when the contact discontinuity reaches $b \approx 50^{\circ}$ for each case.}
\label{fig:oxy-ratio}
\end{figure*}
\section{Results}
\label{sec:results}
Figure \ref{fig:rhot} shows the evolution of density and temperature for the AGNW and SFW models. Within the opening angle, they show a typical structure of the wind blown bubble containing free wind, shocked wind and shocked halo gas \citep{weaver1977}. Note that, the free wind region in case of AGNW is very small and not visible in the density plot because of the colour bar. In a typical wind scenario, the reverse shock appears when the wind ram pressure balances the shocked halo pressure. In a spherically symmetric case, the reverse shock position can be written as (see equation 12 of \cite{weaver1977})
\begin{equation}
\label{eq:r_rs}
r_{\rm rs} \propto \mathcal{L}^{3/10}\,\rho_0^{-3/10}\,v_w^{-1/2}\, t^{2/5} \,\,,
\end{equation}
where $\mathcal{L}$ is the mechanical luminosity and $v_w$ is the free wind velocity, $\rho_0$ is the constant background density and $t$ is the time. Here, we have assumed the mass injection rate, $\dot{M}\, = 2\mathcal{L}/v_w^2$. It is, therefore, clear that the reverse shock in AGNW ($v_w = 0.05\,c$) will be much closer to the centre compared to the SFW for the same luminosity.
\begin{figure*}
\centering
\includegraphics[width=0.8\textheight]{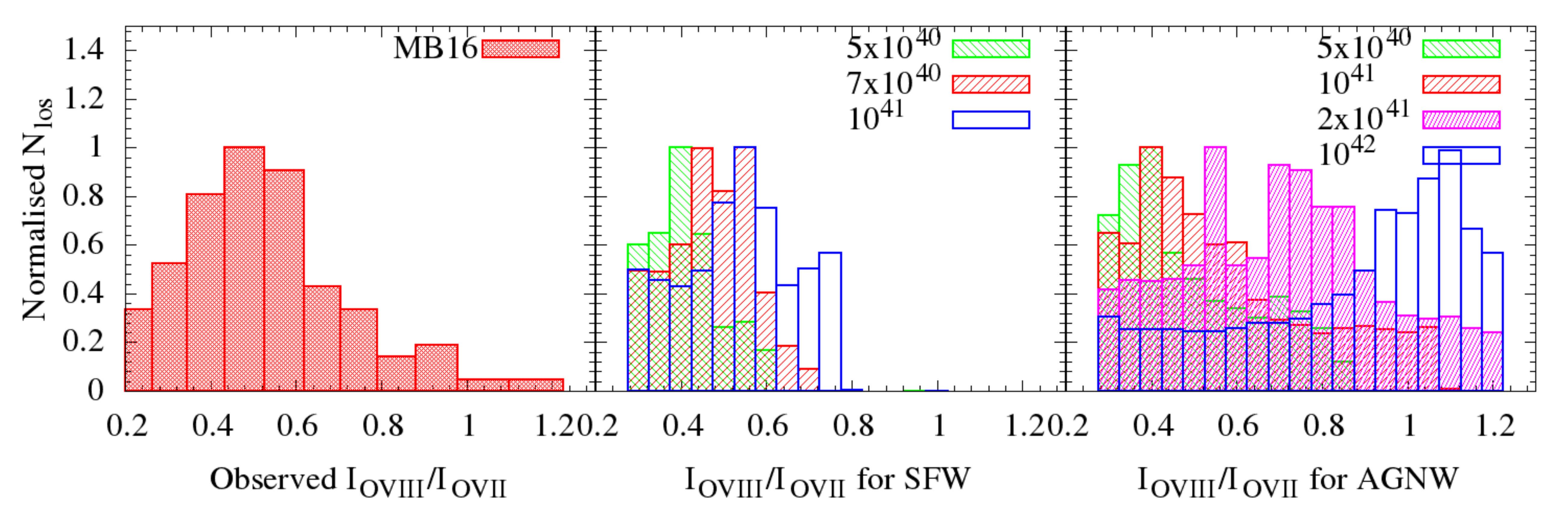}
\caption{Normalised histograms of the OVIII to OVII line ratio obtained at $t=t_{\rm fb}$ as mentioned in Figure \ref{fig:oxy-ratio}. The histogram for $\mathcal{L} = 2\times 10^{41}$ \ergps in AGNW case was obtained at $t=15$ Myrs. The observed values of MB16 are shown in the left panel. The middle and the right panels show the obtained line ratio histograms for the SFW and AGNW cases, respectively. Different mechanical luminosities (given in \ergps) are represented by different box styles. All the ${\rm N}_{\rm los}$ values are normalised with respect to the maximum number of LOSs obtained in corresponding mechanical luminosities. Note that the histograms of intensity ratios are similar for a similar mechanical luminosity, irrespective of whether energy is put via stellar or AGN feedback} 
\label{fig:hist}
\end{figure*}

One important difference between the AGNW and SFW scenarios is the temperature of the shocked wind. For AGNW, it is much higher ($T_{\rm sw} \gtrsim 10^8$K) compared to the SFW ($T_{\rm sw} \sim 10^7$K), and the density in AGNW case is much lower ($\rho_{\rm sw} \sim 10^{-5}$ \mpcc) compared to the SFW case ($\rho_{\rm sw} \sim 10^{-3}$ \mpcc). This is because of the following reasons. Assuming that the total energy is released in the form of kinetic energy, the density of the free wind at any radius, $r$, can be given as 
\begin{equation}
\label{eq:wind-density}
\rho_{\rm w} = 2\mathcal{L}/(\Omega r^2v_{w}^3) \,,
\end{equation}
where, $\Omega$ is the wind opening solid angle. The pressure and temperature of the reverse shocked gas are, therefore,  given as $P_{\rm sw} \propto \rho_{w}v_{w}^2 = 2\mathcal{L}/(\Omega r^2v_{w})$  and $T_{\rm sw} \propto v_w^2$, respectively. This means that a high velocity wind will always create a higher temperature and low density shocked wind. 

It is clear from the above arguments that knowledge of the density structure inside the FBs will help to distinguish between the AGNW and SFW cases. In fact, the best fit model of MB16 (their figure 10a) prefers a high density ($\sim 10^{-3}$ \mpcc) interior of the FBs, which is possible if either SF luminosity is $\sim 10^{41}$ \ergps (SFR $\sim 1$ \mpy) or AGNW luminosity is $\sim 5\times 10^{44}$ \ergps ($\approx 0.8 \mathcal{L}_{\rm edd}$, for a black hole mass of $4\times 10^6$ M$_\odot$), assuming $r_{\rm rs} \sim 2$ kpc and $\Omega = 2\pi$ in equation \ref{eq:wind-density}. \textit{This is a consequence of the fact that the AGNWs are much less mass loaded compared to the SFWs.}

In AGNW case, it is hard to produce the fitted OVIII volume emissivity (shown in figure 10d of MB16) inside the bubble since the emissivity of OVIII lines at $\gtrsim 3\times 10^8$ K is $ n^2 \varepsilon(T) \sim 10^{-28}$ photons s$^{-1}$ cm$^{-3}$, assuming $n \sim 10^{-5}$ cm$^{-3}$ and $\varepsilon(3\times 10^8 K) \sim 2\times 10^{-18}$ photons s$^{-1}$ cm$^{3}$  (see figure \ref{fig:oxy-lines}).  This value is clearly $\sim 7$ orders of magnitude lower than the fitted one. Conduction can, in principle, increase the density inside the bubble and reduce the temperature. Tests with 1D simulations including conduction (see section \ref{subsec:conduction}) show that the temperature of the bubble (i.e. inside the contact discontinuity, which in this case is at $\approx 5$ kpc) is $\gtrsim 10^7$ K and the density is $\lesssim 5\times 10^{-4}$ \mpcc. Therefore, the volume emissivity can increase to $2\times 10^{-24}$ photons s$^{-1}$ cm$^{-3}$, which is still $\sim 3$ orders of magnitude lower than the fitted value $\sim 10^{-21}$ photons s$^{-1}$ cm$^{-3}$. However, we should keep in mind that estimating the emissivity inside the low density bubble is a complex process as it may be contaminated by the shell emission and may not be distinguishable by a simple fitting of a bubble and a shell.

On the other hand, for SFW, the bubbles density $n \sim 10^{-3}$ cm$^{-3}$  and the bubble temperature is $\sim 10^6-10^7$ K, for which the OVIII volume emissivity is $\sim 4\times 10^{-22}$ photons s$^{-1}$ cm$^{-3}$ (assuming $\varepsilon = 10^{-15}$ photons s$^{-1}$ cm$^{3}$) which is much closer to the fitted value. 

In case of the intensity ratio between  OVIII and OVII lines, the comparison becomes non-trivial as the LOS may consist of gas at different temperatures and therefore can have different line ratios compared to a single temperature medium. For direct comparison with the observations, it is necessary to  consider the effects of any intervening or background medium. We, therefore, use our projection software PASS to produce the line intensity maps including the effects of the local bubble and the halo medium extending till $\sim 250$ kpc as explained in section \ref{subsec:pass}. 

While producing line emission maps, it is necessary to know the age of the FBs because the forward shock velocity and hence the shocked halo temperature depends on time as
\begin{equation}
v_{\rm fs} \sim \left( \frac{\mathcal{L}}{\rho_0} \right)^{1/5} t^{-2/5}\,,
\end{equation}
where, the symbols have same meanings as in equation \ref{eq:r_rs}. Therefore, it is necessary to know the region where the gamma-rays are produced. It could either be the forward shock or the reverse shock \citep{lacki2014} or the contact discontinuity \citep{mou2015, crocker2014b} or the region within the contact discontinuity \citep{mertsch2011, sarkar2015b}. Here, we follow \cite{sarkar2015b} and assume that the gamma-rays originate from the region within the contact discontinuity. Therefore, we, set the age of FBs when the contact discontinuity reaches latitude $b \approx 50^\circ$ (height of the FBs). Since the forward shock radius $r_{\rm fs} \sim (\mathcal{L}t^3/\rho_0)^{1/5}$, this age of the FBs is different for different luminosities and is shown in the corresponding panels in figure \ref{fig:oxy-ratio}. 

Figure \ref{fig:oxy-ratio} shows the OVIII to OVII line ratio maps for AGNW (top panel) and SFW (bottom panel) models obtained at the age of the FBs (as explained above) for different luminosities. It shows that the line ratio is highest on the top of the bubble where the shock is the strongest and becomes lower on the either sides where the shock is weaker, a typical behaviour for a bow shock.  Notice that in some cases the emission forms a shell-like feature, this is because the low density interior does not contribute much to the line ratio and most of the emission comes from the shell-like shocked halo gas. 

For a better comparison with the observed data, in Figure \ref{fig:hist}, we also plot histograms of the OVIII to OVII line ratios for different mechanical luminosities and injection scenarios. In this figure we intentionally excluded all the LOSs that have line ratio less than $0.3$ to avoid contamination from the halo gas. 

It is clear from the above figures that only $\mathcal{L} \approx 7\times 10^{40}$ \ergps in case of SFW and $\mathcal{L} \approx 10^{41}$ \ergps in case of AGNW match the observed line ratio. A higher (lower) luminosity in either case produces a line ratio that is more (less) than the observed ones. We, therefore, can constrain the mechanical luminosity of the source of the FBs to be $\mathcal{L} \approx 7\times 10^{40}$ \ergps for the star formation scenario and $\mathcal{L} \approx 10^{41}$ \ergps for the Accretion wind scenario. Note that both the peak and the cut-off of the histograms are characteristic of the injected luminosity rather than just the peak.

The post shock temperature for the SF case corresponds to $\sim 3 \times 10^6$ K, whereas, for the AGNW case, the temperature is $\sim 5\times 10^6$ K at $\theta = 7^\circ$ and falling rapidly to $\sim 3\times 10^6$ K at an angle of $45^\circ$ from the rotation axis. This estimate of temperature is consistent with the measurements by \cite{kataoka2013, gu2016} at the NPS. The similarity of the NPS temperature to the other parts of the FBs is another \textit{dramatic coincidence} that has to be explained if the NPS is not related to the FBs. 
\begin{figure*}
\centering
\includegraphics[width=0.75\textheight]{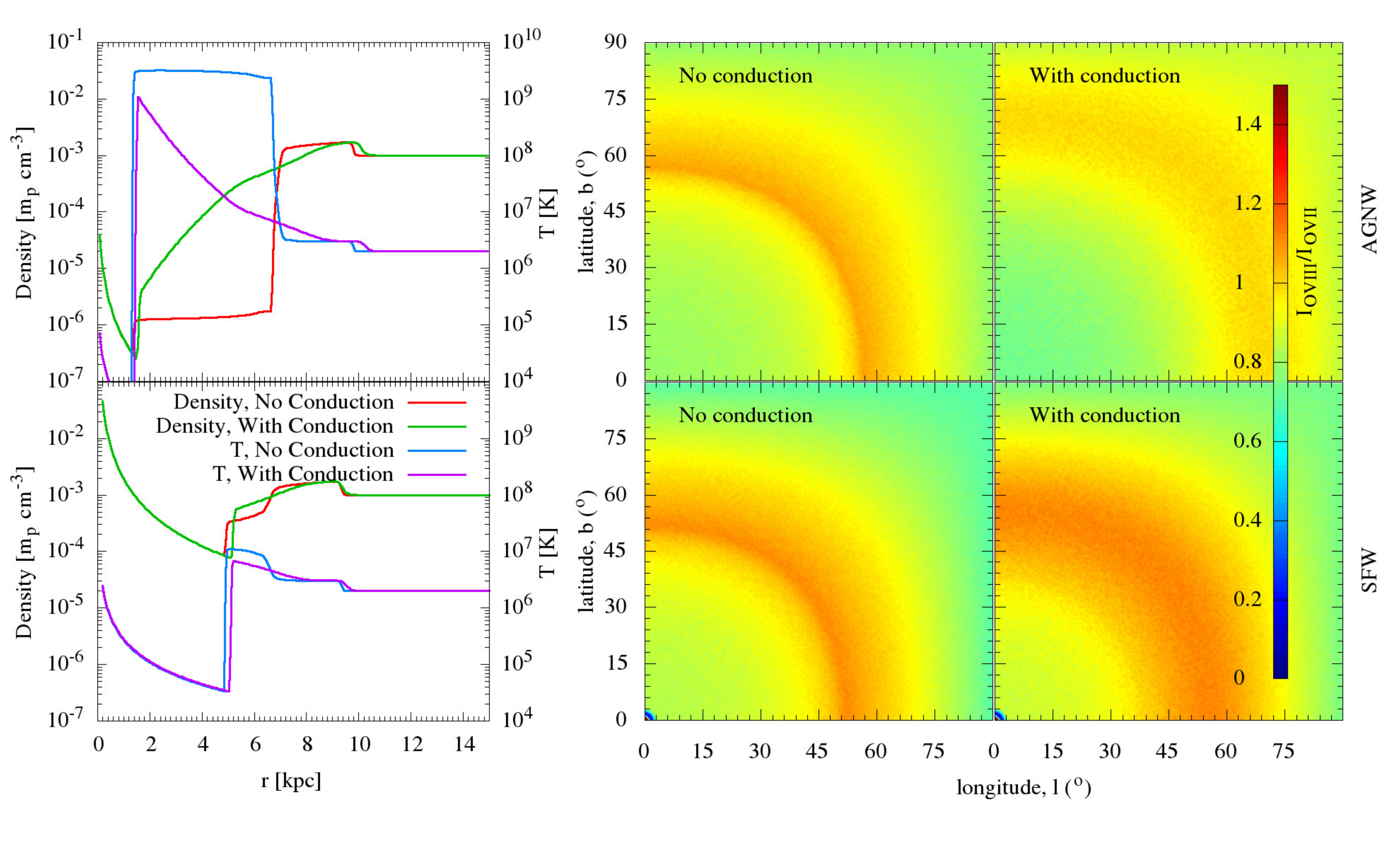}
\caption{Effects of conduction for one dimensional test runs of AGNW (top panel) and SFW (bottom panel) having mechanical luminosity of $\mathcal{L} = 2\times 10^{41}$ \ergps at $20$ Myr. This time-scale has been chosen such that the contact discontinuity reaches longitude $\approx 50^\circ$ when projected. The one dimensional density (left axis) and temperature (right axis) profiles for runs with/without conduction have been shown in the left panel of the figure. The corresponding effects on the OVIII to OVII line ratios have been shown in the middle and right panels. The colour represents the OVIII to OVII line intensity ratio. While calculating the line intensity ratio, we kept the box only till $15$ kpc to avoid contribution from the background halo gas.} 
\label{fig:conduction}
\end{figure*}
\section{Discussion}
\label{sec:discussion}
\subsection{Effects of cosmic ray and magnetic pressure}
\label{subsec:cr_pressure}
 So far in our simulations we have not considered cosmic ray or magnetic pressure on the dynamics of the gas. However, it has been shown that the cosmic ray pressure and the magnetic pressure can contribute approximately half of the thermal energy of the Galactic wind \citep{sarkar2015b}. Therefore, the required energy solely from star formation process to drive the FBs is $\sim 5\times 10^{40}$ \ergps, which corresponds to $\sim 0.5$ \mpy consistent with the estimates by \cite{sarkar2015b}. The estimated mechanical luminosity required only from a SFW wind is, however, dependent on the fraction of the thermal energy in CRs and in the magnetic field. 

\subsection{Enhanced emission beyond FBs}
\label{subsec:enhanced_emission}
One point to notice in Figure \ref{fig:oxy-ratio} is that the line ratio is enhanced beyond the edge of the FBs (extended till $50^\circ$ in latitude and $\sim 20^\circ$ in longitude). Interestingly, such an extended emission (till $\sim 60^\circ-70^\circ$ in both longitude and latitude) in OVIII intensity and the OVIII to OVII line ratio is also noticed in the observations (see figure 4 and 6 of MB16). We speculate that this extended emission can be an indication of the forward shock travelling through the circumgalacitc medium. 

\subsection{Effects of conduction}
\label{subsec:conduction}
Conduction also can affect the dynamics and the density and temperature profiles of the bubble. We, therefore, use isotropic conduction module given in \textsc{PLUTO}. The heat flux is calculated as
\begin{equation}
\label{eq:conduction}
F = \frac{F_{\rm sat}}{F_{\rm sat}+F_{\rm class}} F_{\rm class}\,,
\end{equation}
where, $F_{\rm class} = 5.6 \times 10^{-7} T^{5/2} \nabla T$ is the classical thermal conduction flux. In cases where the temperature gradient is very large, the above equation also takes care of the saturation effects by including $F_{\rm sat} = 5\phi \rho c_{\rm iso}^3$, where, $\phi=0.3$ and $c_{\rm iso}$ is the isothermal sound speed. The effects of conduction are, therefore, more in the case of AGNW because of the high temperature compared to the SFW case. However, incorporation of the thermal conduction in AGNW runs makes the structure of the outer shock highly elongated along the $\theta=0$ axis and forms a very thin jet like feature. In reality, conduction can get suppressed due to the presence of magnetic field. A proper treatment will require incorporation of anisotropic conduction which is beyond the scope of this work. Instead, we present  one-dimensional test runs with/without conduction to understand the effects of conduction. However, we alert the reader that these runs should be taken only as indicative of the actual situation. Moreover, electrons and protons may not have the same temperature behind the shock, as the Coulomb interaction time-scale between these two species can be long, as discussed in section \ref{subsec:e-i}. This may also suppress the thermal conduction.

Figure \ref{fig:conduction} shows the 1D runs with/without conduction. The upper panel shows the runs for AGNW and the lower panel shows the runs for SFW for a mechanical luminosity $=2\times 10^{41}$ \ergps at 20 Myr. The density and temperature profiles for the SFW case show little variation if conduction is present. The effects are large in case of AGNW because of the large temperature inside the bubble. Although the outer shock structure remains almost same, the structure of the density and temperature inside the contact discontinuity ($\approx 5$ kpc) changes by almost two orders of magnitude. To compare the integrated line intensity ratio, we put these 1D bubbles at the Galactic centre and produce the line intensity maps as shown by the colour contours in middle (without conduction) and right panel (with conduction) of Figure \ref{fig:conduction}. The contours show little variation in the line intensity ratio even if conduction is included. This is because the contribution to the line intensity mainly comes from the outer shock which remains almost unaffected by the conduction. Therefore, the line intensity maps presented in Figure \ref{fig:oxy-ratio} are likely to be unaffected by conduction.

One aspect, however, immediately improves in case of AGNW is the OVIII emissivity inside the contact discontinuity. As mentioned earlier, conduction can increase the OVIII emissivity in this case to $\sim 2\times 10^{-24}$ photons s$^{-1}$ cm$^{-3}$.  The exact value however depends on the definition of the bubble i.e. the region where the gamma-ray emission is generated. This will involve careful modelling of the diffusion of CR particles in this scenario. 

Also, notice that the effects of conduction have been overestimated in these simulations. Presence of the magnetic field will decrease the effects of conduction. However, a simple estimate of the average OVIII emissivity in the region within the outer shock shows that the average emissivity is $\sim 4\times 10^{-21}$ photons s$^{-1}$ cm$^{-3}$ in all the cases i.e with/without conduction in AGNW/SFW cases. Since we are looking at projected emission, even the bubble region emits significantly in OVIII because of the outer shock along the sightline. Therefore, it is difficult to distinguish between AGNW and SFW using the OVIII emissivity fitted by MB16.

\subsection{Electron-proton energy equilibration}
\label{subsec:e-i}

Eq. \ref{eq:conduction} assumes that electrons and protons have the same temperature. This assumption is valid only when the electron-proton energy exchange time due to Coulomb collisions is short enough compared to the dynamical time. This time-scale is (using Eq. 5-31 in \citealt{spitzer1956}), 
\begin{equation}
\label{eq:t_eq}
t_{eq} \sim 0.25 {\rm ~Myr}~ T_6^{3/2}/n_{-3}\,\,,
\end{equation}
where, $T_6$ is the electron temperature in the units of $10^6$ K and $n_{-3}$ is the proton/electron number density in units of $10^{-3}$ cm$^{-3}$. The corresponding length scale required to attain equilibrium is $l_{eq} = v t_{eq} \sim 75\, T_6^{3/2}/n_{-3}$ pc for $v=300$ km s$^{-1}$ ($v$ is the flow velocity).  Thus, for the outer shock density and temperature (for both SFW and AGNW scenarios considered here) $t_{eq}$ is shorter than the age of FBs, and therefore the electron and proton temperature behind the outer shock can be treated as equal.
 
 For the much stronger reverse shock in the AGNW scenario (top-left panel of Fig. \ref{fig:conduction}), the post-shock temperature is  $\sim 10^9$ K and density is $\sim 10^{-6}$ cm$^{-3}$. The electron-proton energy exchange time for these parameters is $\sim 10^7$ Myr ! For the SFW scenario (top-left panel of Fig. \ref{fig:conduction}) $t_{eq}$ is $\sim 15$ Myr, and even here the assumption of equal electron and proton temperature is only marginally valid. Thus the strong reverse shock is in the collisionless regime, and the electron temperature is expected to be much smaller than the proton temperature (e.g., see Fig. 2 in \citealt{ghavamian2007}). Therefore, the effects of thermal conduction are exaggerated in the top panels of Figure \ref{fig:conduction}, and in reality the density in the bubble (particularly for the AGNW scenario) may be closer to the case without conduction.

For a strong outer shock ($M \sim 10$, representative of a high luminosity wind) the forward shock temperature can become $T\sim 10^8$ K for which $t_{\rm eq}\sim 100$ Myr (Eq. \ref{eq:t_eq}). In such a case, the electrons are expected to be much cooler than the protons and, therefore, the outer shock strength may be underestimated by the OVIII/OVII ratio. We can estimate the maximum luminosity for which our analysis, which hinges on equal electron and proton temperature, of the outer shock strength is valid.  The outer shock temperature for a mechanical luminosity $\mathcal{L}$ can be approximated in a spherically symmetric and constant background density case as
\begin{equation}
T_{\rm shock} \approx 2.5\times 10^7  \mathcal{L}_{42}^{2/5}\,n_{-3}^{-2/5}\,t_{\rm dyn, Myr}^{-4/5}\,\,\,\, \mbox{K} \,,
\end{equation}
where, $ \mathcal{L} = 10^{42}\mathcal{L}_{42}$ \ergps and $t_{\rm dyn, Myr}$ is the time in units of Myr which is given by
\begin{equation}
\label{eq:t_dyn}
t_{\rm dyn, Myr} = 11\, R_{\rm 10kpc}^{5/3}\,n_{-3}^{1/3}\,\mathcal{L}_{42}^{-1/3}\,.
\end{equation}
Here, $R = 10 R_{\rm 10kpc}$ kpc is the outer shock radius. Therefore, we can write Eq. \ref{eq:t_eq} as
\begin{equation}
t_{\rm eq, Myr} \sim 1.74\, \mathcal{L}_{42}\,n_{-3}^{-2}\, R_{\rm 10kpc}^{-2}\,.
\end{equation}
Now, for the electron and proton temperature to be equal, $t_{\rm eq} \lesssim t_{\rm dyn}$, which means
\begin{equation}
\label{eq:max_L}
\mathcal{L} \lesssim 4\times 10^{42}\, n_{-3}^{7/4}\,R_{\rm 10kpc}^{11/4}\,\,\,\,  \mbox{\ergps}\,.
\end{equation}
Therefore, our analysis of the outer shock strength is valid for $\mathcal{L} \lesssim 4 \times 10^{42}$ \ergps . Note that the above calculation assumes that the shock is expanding in a constant density medium. In reality, the shock expands in a stratified CGM for which an analytical solution in general is difficult to obtain. We can obtain an upper limit on $\mathcal{L}$ by plugging in the lowest plausible value for $n$ ($\sim 5\times 10^{-4}$ cm$^{-3}$; the minimum CGM density within $10$ kpc; see Fig. \ref{fig:rho}) in Eq. \ref{eq:max_L}. This assures that the assumption of electron-proton equilibrium at the outer shock is definitely valid for $\mathcal{L} \lesssim 10^{42}$ \ergps.

For higher mechanical luminosities, the electron temperature ($T_e$) can be lower than the equilibrium shock temperature ($T_{\rm shock}$) obtained from shock jump conditions. However, $T_e$ at $t = t_{\rm dyn}$ (time at which the outer shock reaches the observed size of the X-ray shell) is still higher than the electron temperatures corresponding to low luminosity cases (see appendix \ref{app:observable_Te} for details). Therefore, any luminosity more than $10^{42}$ \ergps would yield electron temperature higher than the electron temperature of $\mathcal{L} = 10^{42}$ \ergps case and would be observable in the OVIII/OVIII ratio map. Therefore,  a weak outer shock strength is the only possible solution for explaining the observed OVIII to OVII line ratio.  
\subsection{AGNW vs. SFW}
\label{subsec:agn_vs_sfw}
In case of a SFW, the obtained mechanical luminosity ($5\times 10^{40}$ \ergps) corresponds to a SFR$\sim 0.5$ \mpy (see equation \ref{eq:mech-L} and section \ref{subsec:cr_pressure}). Notice that this value is slightly larger compared to the observations by \cite{yusuf-zadeh2009}, who found SFR $\sim 0.1$ \mpy. However, recent discovery of a $\sim 100$ pc molecular ring can, in principle, host a higher SFR. Also, note that the required rate of SFR depends on the exact amount of CR and magnetic energy density inside the bubbles. 

On the other hand,  linear polarisation of $\gtrsim 150$ GHz emission, and  IR and X-ray variability of the Sgr A$^*$ suggests that the current accretion rate of the GCBH is $\sim 10^{-9}\,\hbox{-}\,10^{-7}$ \mpy \citep{quataert2000, agol2000, yuan2003, marrone2006}, which corresponds to a mechanical luminosity of $\sim 5\times 10^{36\hbox{-}38}$ \ergps, assuming an efficiency factor of $0.1$ (see section \ref{sec:intro}). However, in order to explain the X-ray luminosity around the Sgr A$^*$, it has been suggested that the past accretion rate of the GCBH could have been $10^{3\hbox{--}4}$ higher than the present day accretion rate \citep{totani2006}. This means that the GCBH mechanical luminosity was $\lesssim 5\times 10^{39\hbox{--}41}$ \ergps. Although there is a large uncertainty in the past mechanical luminosity, it is surprisingly close the required rate of $10^{41}$ \ergps. Also, we must note that the black hole activity is highly variable in time and it is the average mechanical luminosity that should be considered.

In this paper, though we constrain the mechanical luminosity for the source driving the FBs, the degeneracy between the SFW and the AGNW models still remains. One way to distinguish between these two models is probably the kinematics of the hot gas inside the bubbles (i.e. inside the contact discontinuity). As noted by \cite{sarkar2015b}, the velocity range of the hot wind for SFW can be $\sim \pm 600$ \kmps . However, in case of AGNW, This velocity range will be much higher. Another way is to measure the temperature along the outer edge of the FBs. In AGNW case, the outer shock is relatively more anisotropic than the SFW case. This is because the AGNW is completely kinetic energy driven and has a large velocity anisotropy perpendicular to the disc thus producing a strong bow shock and, therefore, producing a somewhat anisotropic shock temperature. The SFW, on the other hand, has a large fraction of energy in the form of internal energy and hence the outer shock structure is more isotropic (see Figure \ref{fig:rhot}). However, one must note that measuring the temperature along the edge of the FBs using the OVIII to OVII line ratio is likely to be contaminated by the detailed structure of the MW halo gas as the contribution from the background halo gas is non-negligible.

\section{Summary}
We have explored different driving mechanisms to inflate the FBs: one, a central black hole driven wind (AGNW), and second, a star formation driven wind (SFW). The winds have been launched at the Galactic centre and have been allowed to propagate through a realistic distribution of the MW halo gas. We compare our numerical simulations of SNe and AGN wind models with the best fit model of \cite{miller2016}.  We find that irrespective of the driving mechanism - AGNW or SFW, the total luminosity required to produce the observed OVIII to OVII line ratio is $\approx 0.7-1\times 10^{41}$ \ergps.  The given luminosity also constrains the age of the FBs to be $\sim 20$ Myr.

The shocked halo temperature is estimated to be $\approx 3\times 10^6$ K in most of the forward shock.  For a weak shock travelling through a $2\times 10^6$ K halo gas, this temperature would correspond to a shock velocity of $\sim 300$ \kmps. The corresponding temperature is highly anisotropic in case of a AGNW for which the temperature ranges from $5\times 10^6$ K to $3\times 10^6$ K. These values are slightly lower compared to the estimates by \cite{miller2016} who found the temperature to be $\approx 5\times 10^6$K based on the same data. Our temperature estimate is, however, consistent with the temperature measurements by \cite{kataoka2013} and \cite{gu2016} at the NPS, which indicates that the NPS has likely originated from the same activity that gave rise to the FBs. 

\section*{Acknowledgements}
It is a pleasure to thank David Eichler for helpful discussions on the electron-proton equilibration temperature. We thank Yoshiaki Sofue and Tomonori Totani for useful comments on the draft. KCS thanks Saurabh Singh for helps in debugging PASS. We also thank the anonymous referee for constructive comments that helped improve this paper. This work is partly supported by an India-Israel joint research grant (6- 10/2014[IC]).


%
%
%
%
%
\appendix

\section{Injection radius for SNe driven winds}
\label{app:injection_radii}
\begin{figure*}
\centering
\includegraphics[width=0.75\textheight]{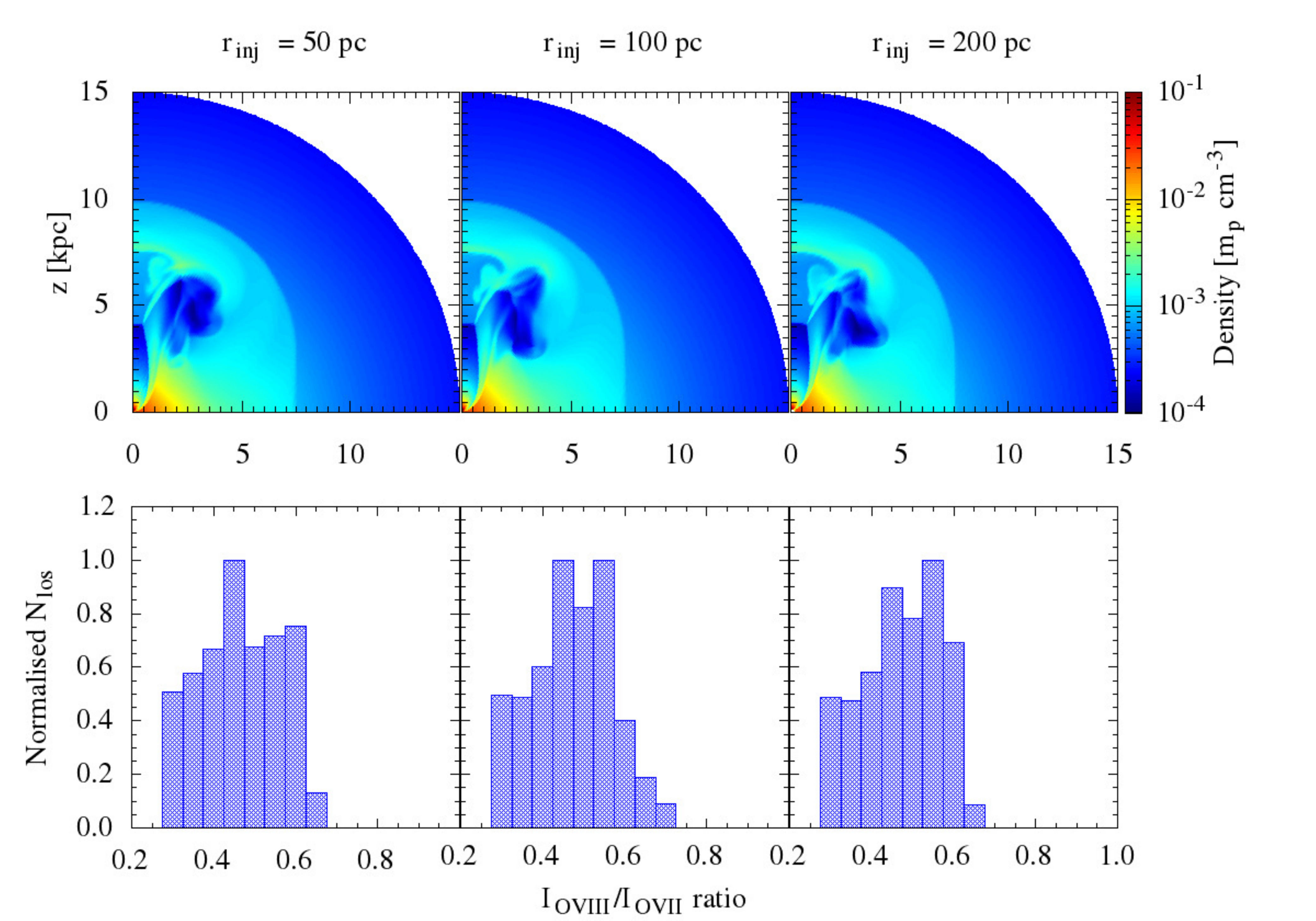}
\caption{Effects of different injection radius for SFW cases. The top panel shows the density contours for $\mathcal{L} = 7\times 10^{40}$ \ergps at $t = 25$ Myr and the bottom panel shows the corresponding histograms of OVIII to OVII line ratios.} 
\label{fig:rinj_comp}
\end{figure*}
As mentioned in the main text, we have chosen the injection radius (which is the same as the radius of the inner boundary) for the SFW to be at $r_{\rm inj} = 100$ pc which is also assumed to be the sonic radius of the wind. Though this particular choice of the radius is somewhat arbitrary, any deviation from it does not affect the results. Figure \ref{fig:rinj_comp} shows the density contours and the corresponding histograms of OVIII to OVII line ratios for $\mathcal{L} = 7\times 10^{40}$ \ergps at $t = 25$ Myr but for injection radii of $50$, $100$ and $200$ pc. Other than some tiny details, the results are consistent with each other.
\section{Electron temperature for high luminosity winds}
\label{app:observable_Te}
\begin{figure*}
\includegraphics[width=0.32\textheight, angle=-90, trim={2cm, 3.5cm, 0cm, 4.0cm}, clip=true]{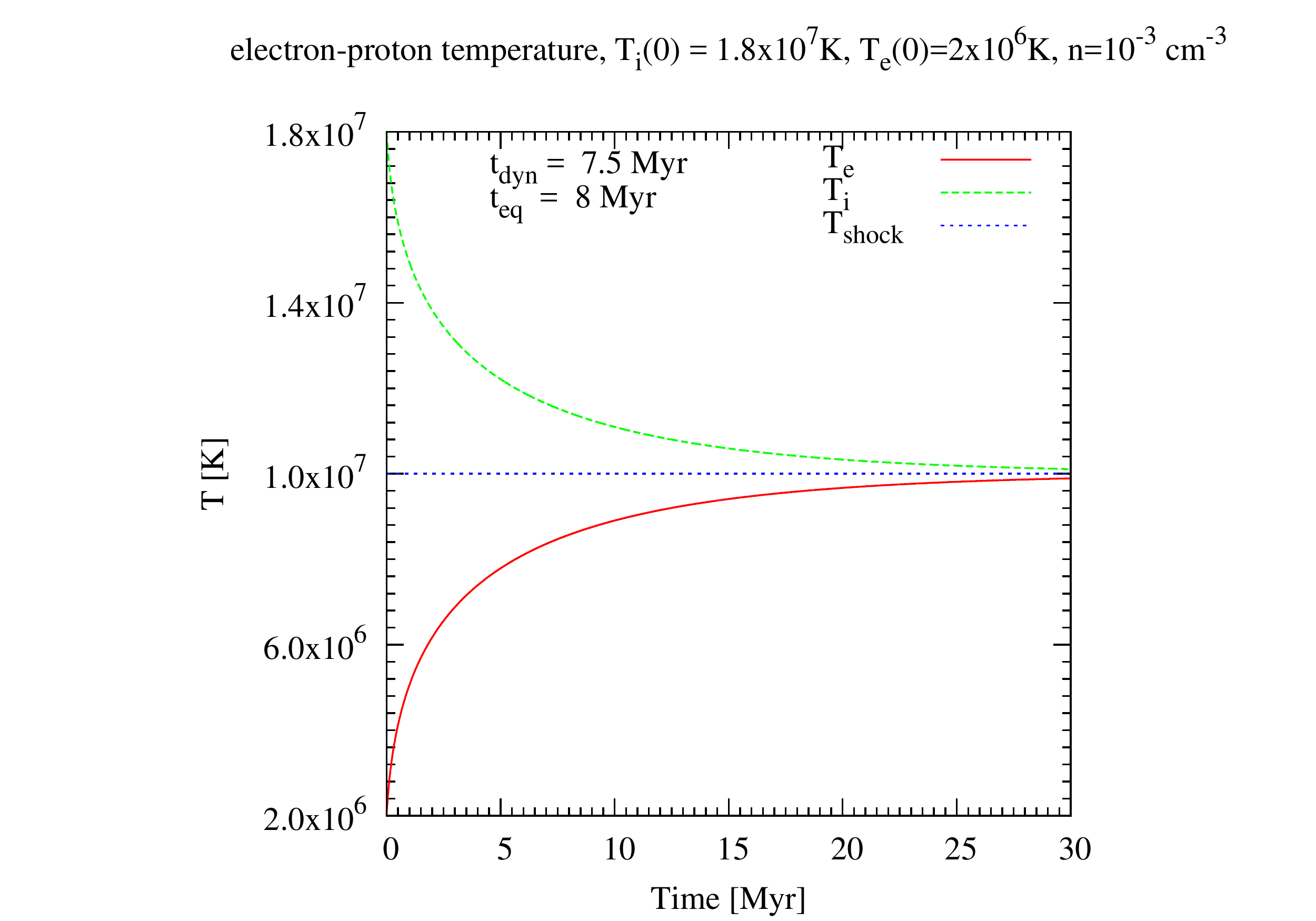}
\includegraphics[width=0.33\textheight, angle=-90, trim={1.5cm, 4.0cm, 0cm, 4.0cm}, clip=true]{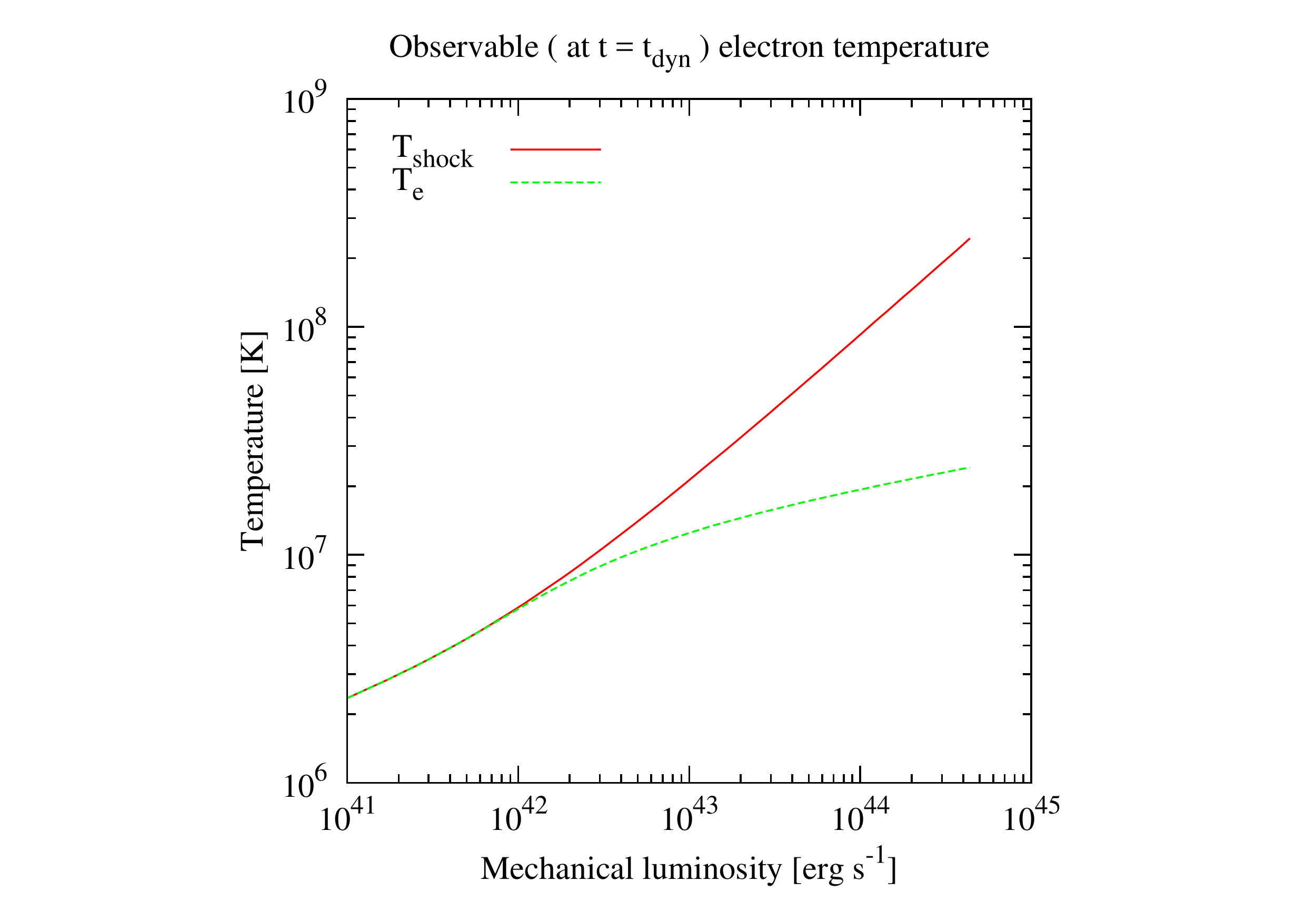}
\caption{Left panel: Evolution of the electron and proton temperatures (Eqs. \ref{eq:Te}, \ref{eq:Ti}) towards an equilibrium shock temperature of $10^7$ K corresponding to $\mathcal{L} \approx 3\times 10^{42}$ \ergps. Right panel: Electron temperature at $t = t_{\rm dyn}$ (time appropriate for FB observations; see Eq. \ref{eq:t_dyn}) compared to the equilibrium shock temperature ($T_{\rm shock} \equiv [T_e + T_i]/2$) for different luminosities shows a monotonic increase (although slower than $T_{\rm shock}$) in $T_e$ with an increasing luminosity even in the collisionless regime. Both the figures are obtained for $n = 10^{-3}$ cm$^{-3}$.} 
\label{fig:Te_evolution}
\end{figure*}
To study the evolution of electron temperature $T_e$ and ion temperature $T_i$, we assume the simplest picture that these two species exchange energy only via Coulomb collisions, that there is no relative bulk velocity between them and that the electron and ion number densities are equal, i.e. $n_i \approx n_e \equiv n$. The energy equations for these two species can then be approximated  as \citep{braginskii1965}
\begin{eqnarray}
\label{eq:Te}
n_e k_B \left ( \frac{3}{2} \frac{d T_e}{dt} +  T_e  {\bf \nabla \cdot v} \right ) &=& - 3\frac{m_e}{m_p} \frac{n k_B}{\tau_e} \left( T_e - T_i\right),  \\ 
\label{eq:Ti}
n_i k_B \left ( \frac{3}{2} \frac{d T_i}{dt} +  T_i  {\bf \nabla \cdot v} \right )  &=&  3\frac{m_e}{m_p} \frac{n k_B}{\tau_e} \left( T_e - T_i\right), \, 
\end{eqnarray}
where ${\bf v}$ is the bulk velocity, $\tau_e = 3.44\times 10^5 \frac{\left(k_B T_e/{\rm eV}\right)^{3/2}}{n \lambda}$ sec and $\lambda \approx 15$ is the Coulomb logarithm.  Other constants have their usual meanings. Setting the compression term to zero (${\bf \nabla \cdot v} = 0$ in Eqs. \ref{eq:Te} \& \ref{eq:Ti}), we can solve for the electron and ion temperatures of the post-shock gas. We assume that $T_i (t=0) = T_{i,2}$ and $T_e (t=0) = T_{e,2}$, the ion and electron post shock temperature respectively. Assuming that both electron and ion bulk kinetic energies are thermalised independently at shocks (observations suggest that electrons are heated more than this estimate so we can treat our $T_e$ as a conservative lower limit on the electron temperature), the shock energy gets distributed among electrons and protons according to their mass and the Mach number of the shock. Therefore, the post shock electron and ion temperatures can be written, respectively, as (see Eqs. 19, 21 in \citealt{vink2015})
\begin{eqnarray}
\label{eq:Te2}
T_{e,2} &=& T_{e,1}\chi^{\gamma-1} + \frac{\mu_e m_p v_s^2}{2k_B} \left( \frac{\gamma-1}{\gamma}\right)\left(1-\frac{1}{\chi^2} \right), \\ 
\label{eq:Ti2}
T_{i,2} &=& T_{i,1} \left(2-\chi^{\gamma-1}\right) + \frac{\mu_i m_p v_s^2}{2k_B} \left( \frac{\gamma-1}{\gamma}\right)\left(1-\frac{1}{\chi^2} \right) \,,
\end{eqnarray}
where, $T_{e,1} = T_{i,1} = 2\times 10^6$ K is the pre-shocked halo temperature, $\chi = \left(\gamma + 1\right)M^2/\left((\gamma-1)M^2+2 \right)$ is the density jump behind the shock for a Mach number of $M$, $v_s$ is the shock velocity, $\mu_e = m_e/m_p$ and $\mu_i = 1.27$ for Solar metallicity.

Figure \ref{fig:Te_evolution} shows the evolution of the electron and ion temperatures (Eqs. \ref{eq:Te} \& \ref{eq:Ti} with $\nabla \cdot {\bf v}=0$ and initial conditions given by Eqs. \ref{eq:Te2} \& \ref{eq:Ti2}) behind the outer shock. The left panel of the figure shows that although it takes few$\times t_{\rm eq}$ (Eq. \ref{eq:t_eq}) to come to equilibrium ($T_e \approx T_i$), initially $T_e$ rises very sharply and attains a value $\approx 90\%$ of $T_{\rm shock} \equiv (T_e+T_i)/2$ within an equilibrium time $t_{\rm eq}$. For comparison with FB observations the electron temperature behind the shock should be evaluated at $t = t_{\rm dyn}$ (see Eq. \ref{eq:t_dyn}). We plot this $T_e (t= t_{\rm dyn})$ in the right panel of figure \ref{fig:Te_evolution}. We notice that although current $T_e$ lags the equilibrium shock temperature, it still increases monotonically with the mechanical luminosity. Therefore, even if $t_{\rm eq} > t_{\rm dyn}$ for high luminosity cases (i.e. the shock is collisionless for ${\cal L} \gtrsim 10^{42}$ erg s$^{-1}$), the electron temperature is still too high to explain the observed OVIII/OVII line ratio.

\end{document}